\documentclass[aps,prl,twocolumn,showpacs,amsmath,amssymb]{revtex4}
\usepackage{graphics}
\usepackage{epsfig}
\begin{document}
\title{\Large \bf{Quantum pumping: Coherent Rings versus Open Conductors}} 
\author{M. Moskalets$^{1}$ and M. B\"uttiker$^2$}
\affiliation{
         $^1$Department of Metal and Semiconductor Physics,\\
        National Technical University "Kharkov Polytechnical Institute",
        61002 Kharkov, Ukraine\\
        $^2$D\'epartement de Physique Th\'eorique, Universit\'e de Gen\`eve,
        CH-1211 Gen\`eve 4, Switzerland\\}

\date\today

\begin{abstract}
We examine adiabatic quantum pumping generated by an oscillating scatterer 
embedded in a one-dimensional ballistic ring and compare it with pumping
caused by the same scatterer connected to external reservoirs. 
The pumped current for an open conductor, paradoxically, is non-zero
even in the limit of vanishing transmission. In contrast, for the 
ring geometry the pumped current vanishes in the limit of vanishing transmission. 
We explain this paradoxical result and demonstrate that the physics underlying 
adiabatic pumping is the same in open and in closed systems.
   \end{abstract}

\pacs{72.10.-d, 73.23.-b, 73.40.Ei}

\maketitle

\small 

\begin{figure}[b]
  \vspace{3mm}
  \centerline{
   \epsfxsize 8cm
   \epsffile{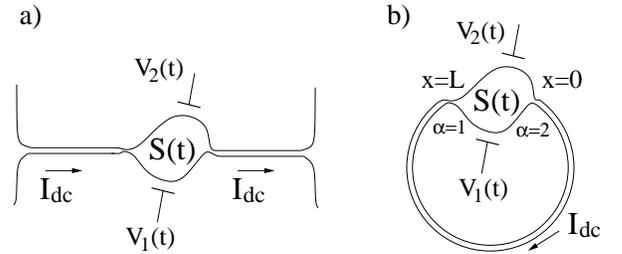}
             }
  \vspace{3mm}
  \nopagebreak
    \caption{
A quantum dot with scattering matrix $\hat S$ and two leads.
Two nearby metallic gates modulate the shape 
and hence the scattering properties of the dot.
If the gate potentials $V_{1}$ and $V_{2}$ change cyclically but shifted in phase then
a current $I_{dc}$ can arise in the leads.
(a) - in an open conductor the current $I_{dc}$ flows between the external reservoirs;
(b) - in a closed conductor the current $I_{dc}$ flows along a ring of length $L$ 
formed by the leads.
The Greek letter $\alpha$ numbers the scattering channels. 
}
\label{fig1}
\end{figure}

Adiabatic particle transport under slow cyclic 
evolution of an internal potential has a long history \cite{Thouless83}. 
However, only recently was such adiabatic transport investigated
experimentally in open phase coherent 
mesoscopic conductors \cite{SMCG99}. This has stimulated increasing interest
in this subject
\cite{
Brouwer98,BTP94,AA98,ZSA99,AK00,SAA00,WWG00,AEGS00,VAA01,
PB01,MM01,MB01,AEGS01,EWAL02,MB02,Cohen02,MBstrong02,SC02,MBhidden,
AEGS03,BDR03}.
The experiment by Switkes et al. ~\cite{SMCG99} was carried out on open samples 
coupled to reservoirs \cite{note} via leads (see Fig.\ref{fig1}a).

In open systems the electron spectrum is continuous and even a slowly  
oscillating scatterer induces transitions between electron states. 
Therefore a purely quantum-mechanical adiabaticity condition 
is always violated. 
However if the oscillation frequency $\omega$ is small compared 
to the inverse time $\tau_{T}^{-1}$ taken for carriers to traverse 
the scatterer, then such a pump can be termed {\it adiabatic}.
Brouwer Ref.~\cite{Brouwer98} gave an elegant formulation of adiabatic 
($\omega\tau_{T}\ll 1$) quantum pumping based on the scattering matrix approach 
to low frequency ac transport in phase coherent mesoscopic 
systems \cite{BTP94}. 

In contrast, in closed systems, when the sample's leads 
are bent back to form a ring (see Fig.\ref{fig1}b),  
the spectrum is discrete. In this case, if the frequency $\omega$ 
is small compared with the level spacing, then the true quantum-mechanical 
adiabaticity condition can be achieved. 
Formally the conditions for the existence of an adiabatic pumped current in 
open and in closed systems are the same: the oscillating scatterer has to break
the time reversal invariance \cite{Brouwer98,MBstrong02,MBhidden}. 

Interestingly, 
we find that the expressions for the pumped current in the open 
and closed cases differ significantly. 
To illustrate this difference we consider a simple specific model: 
A scatterer with two one-channel leads.
In the absence of magnetic fields such a model is described by
the symmetric $2\times 2$ scattering matrix 
\begin{equation}
\label{Eq1}
\hat S = \left(
  \begin{array}{cc}
       \sqrt{R}e^{-i\theta}   & i\sqrt{T}            \\
       i\sqrt{T}             & \sqrt{R}e^{i\theta} \\
  \end{array}
\right).
\end{equation}
\noindent 
Here $R$ and $T$ are the reflection and the transmission 
probability, respectively ($R+T=1$). 
The phase $\theta$ characterizes the asymmetry of particle 
reflection to the left and to the right. 
We assume the quantities $R,T = 1- R ,\theta$ to be functions 
of external parameters varying with frequency $\omega$. 
If the scatterer is connected to the external reservoirs Fig.\ref{fig1}a
then the adiabatically pumped current $I_{dc}$ is \cite{AA98,SAA00}

\begin{equation}
\label{Eq2}
  I_{dc}^{(open)} =  \frac{e\omega}{4\pi^2}
\int\limits_{0}^{\cal T} dt R 
 \frac{\partial\theta}{\partial t}.
\end{equation}

\noindent 
Here ${\cal T} = 2\pi/\omega$ is the period of a pumping cycle.
For the closed ring-geometry Fig.\ref{fig1}b, 
we will show below that each energy level 
$E^{(l)}$ can carry a pumped current $I_{dc}^{(l)}$ given by

\begin{equation}
\label{Eq3}
 I_{dc}^{(l)} =  \frac{e\omega}{4\pi} (-1)^l
\int\limits_{0}^{\cal T} dt \sqrt{\frac{T}{R}} 
 \frac{\partial\theta}{\partial t}.
\end{equation}

\noindent The full current circulating in a ring is given 
by the sum over all occupied levels. 
Eq.(\ref{Eq3}) is valid only if $R\neq 0$.

There is a striking difference between Eq.(\ref{Eq2}) and Eq.(\ref{Eq3}): 
Eq.(\ref{Eq2}) predicts pumping even in the limit of R = 1
if only the phase $\theta$ changes by $2\pi$ during a pump cycle. 
This result is paradoxical because at $R=1$ the two reservoirs are in fact
completely decoupled from each other. 
In contrast, for the ring, the expression for the current  Eq.(\ref{Eq3}) seems to be more
reasonable because it gives no pumped current at $R=1$ 
(when the ring is transformed into a wire disconnected from a cavity).

We now first discuss the resolution of this puzzling difference 
and only subsequently discuss the derivation of Eq.(\ref{Eq3}).
To resolve the paradox we analyze the topology 
of adiabatic pump cycles and show that not each cycle is a genuine pump cycle.  
Moreover for a true pump cycle both 
Eq.(\ref{Eq2}) and Eq.(\ref{Eq3}) simultaneously either give a pumped current or
give no pumped current.

From Eq.(\ref{Eq2}) it follows that the charge 
$Q = ({2\pi}/{\omega})I_{dc}^{(open)}$ 
pumped during the ${\rm LPC}$ in the open case is
exactly quantized $Q[{\rm (LPC)}_{n}] = ne$.
In some papers \cite{AA98,SAA00,AK00,MM01,SC02} this, in fact, topological result 
was used to analyze the conditions for quantization of the pumped charge.
However, 
we can ask: How can a charge $ne$ be pumped between reservoirs 
if during the cycle under consideration the reservoirs 
are completely decoupled from each other since $R=1$?

\begin{figure}[t]
  \vspace{3mm}
  \centerline{
   \epsfxsize 8cm
   \epsffile{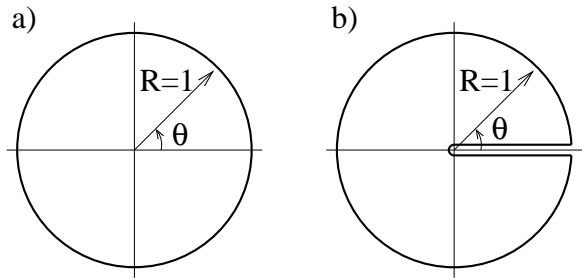}
             }
  \vspace{3mm}
  \nopagebreak
  \caption{
(a)- A limiting pump cycle. 
During the cycle the reflection probability is constant: $R=1$, 
the phase $\theta$ changes by $2\pi$: $0 \leq \theta < 2\pi$,
and a dipole charge $\pm e$ is accumulated inside the scattering region.
(b) - A true limiting pump cycle. 
After the cycle the reflection probability $R$ and 
the phase $\theta$ return to their initial values.
There is no dipole charge accumulation inside the scattering region.
The charge transferred  between the reservoirs is $Q = e$ for both cycles.
}
\label{fig2}
\end{figure}

If the sample is characterized by the scattering matrix Eq.(\ref{Eq1}) 
then any pump cycle can be 
represented by some closed curve in the plane with
$\sqrt{R}$ and $\theta$ being the polar coordinates.
Because the maximum value for $R$ is unity each pump cycle
lies inside the circle of radius $R=1$.
This circle (shown in Fig.\ref{fig2}a) itself represents a pump cycle.
We call this cycle a "limiting pump cycle" (${\rm LPC}$). 
In fact there is a set of cycles which differ from each other
by how many times $n$ the curve encircles the origin. 
We will use this winding number $n$ to distinguish different ${\rm LPC}'s$.
During the ${\rm (LPC)}_{n}$ the parameters of the scattering matrix 
change as follows: $R = 1$, $0 \leq \theta < 2\pi n$.
Note that any pump cycle with $R(t)\leq 1$ characterized 
by the winding number $n$ lies inside the ${\rm (LPC)}_{n}$.

The answer is the following. 
During the ${\rm LPC}$ the charge $ne$ comes from the left reservoir
and accumulates on the left side of the sample. 
In addition the same charge flows from the right side of the sample
to the right reservoir. As a result the charge $ne$ is effectively transferred
between the reservoirs. But this is not only the result of the LPC.
There is an unavoidable (dipole) charge accumulation inside the sample during the LPC.
Formally we can show this as follows. Since 
the direct transmission through the sample is prohibited,
$S_{12}=S_{21}=0$, 
the sample can effectively be viewed as a mesoscopic capacitor \cite{BTP93,PTB96}. 
The left and the right sides of a sample are the plates of a capacitor
which connect to the left and to the right reservoirs, respectively.
We can define the (one-channel) scattering matrices 
$S_L$ and $S_R$ for the left and for the right plates, respectively:
$S_L \equiv S_{11} = e^{-i\theta}$,
$S_R \equiv S_{22} = e^{i\theta}$.
According to the Friedel sum rule \cite{Friedel52} the 
variation of the scattering matrix defines the variation of 
the charge on the scatterer: $\delta Q = \frac{e}{2\pi i} \delta \ln({\rm det}[S])$.
Therefore the charge variation on the plates of a capacitor is  

\begin{equation}
\label{Eq4}
\delta Q_L = - \frac{\delta\theta}{2\pi}e, \,\,\, 
\delta Q_R =  \frac{\delta\theta}{2\pi}e.
\end{equation}
Although formally the scattering matrices $S_L$ and $S_R$
are periodic in $\theta$ with the period of $2\pi$, 
the absolute value of $\theta$ has nevertheless a strict physical meaning:
The change of $\theta$ determines the change of the charge of a capacitor.
Thus we can conclude that after each LPC the sample does not
return to its initial state but rather the sample accumulates
some dipole charge inside: 
$Q_R[{\rm (LPC)}_{n}] = -Q_L[{\rm (LPC)}_{n}] = ne$.
Note that the same amount of charge $ne$ is effectively transferred between 
the reservoirs (during this cycle). Due to the build-up of a dipole charge 
the scatterer cannot operate for an infinitely long 
time and therefore the LPC is not a "true" pump cycle.

To obtain a true pump cycle 
(with no dipole charge accumulation inside the scatterer) 
we have to return the sample to its initial state.
To this end we need to discharge the capacitor. 
Formally this means that during such a process (discharging) 
the parameter $\theta$ has to change from $2\pi n$ to zero
(if the cycle starts with $\theta = 0$). 
Physically this means that we have to make an electrical contact 
between the plates. In other words, the sample has to become
(at least partially) transmitting for a moment.

The discharging can be realized in a number of ways.
For instance, we can transform any ${\rm (LPC)}_{n}$ into 
a "true limiting pump cycle" ${\rm (TLPC)}_{n}$
as shown in Fig.\ref{fig2}b for $n=1$.
In this case the overall pumped charge remains the same $Q = en$.
Importantly, the system now returns to its initial state after
the completion of each pump cycle. Hence the TLPC can be repeated as many times
as desired.

From this discussion one can see that in 
the integral representation Eq.(\ref{Eq2}) generally consists of two parts. 
The first is a true pumped current 
which results from the direct charge exchange between the outside reservoirs. 
The second is a pseudo pumped current
which is a consequence of a 
charge exchange between the scatterer and each of the reservoirs
separately.
Strictly speaking this last part does not follow from the calculations of the pumped current
(see e.g., Ref.~\cite{Brouwer98}) and it arises exclusively due to the representation 
of the pumped current as a contour integral in the scattering matrix space.
To be consistent we can use the integral representation Eq.(\ref{Eq2}) only 
with the restriction that any cycle showing a pseudo pump effect must be excluded.
Thus  the (true) pumped current has no contribution coming
from the topology. This is in agreement with Ref.~\cite{AEGS01}.

We can therefore conclude that for any true pump cycle 
Eqs.(\ref{Eq2}) and (\ref{Eq3}) 
both give either zero or give a pumped current. 
Thus the same scatterer subject to the same (true) pump cycle 
produces current in the open case Fig.\ref{fig1}a as well as in the closed case Fig.\ref{fig1}b.
Therefore the physics responsible for generating a pump effect is the same
in open and in closed geometries. Of course 
because of the different spectra (continuous and discrete) the 
pumped currents in an open and in a closed system 
can be of very different magnitudes.

Now we proceed to the discussion of 
the pumped current in a closed geometry to prove the announced result
Eq.(\ref{Eq3}).
We use the scattering matrix approach 
to pumping in closed systems developed in Ref.~\cite{MBhidden}.
This allows us to consider the pump effect in closed and open cases on the same footing.
To clarify the essential physics of an adiabatic quantum pump effect in closed systems
we consider a simple model: 
A one-dimensional ring of length $L$ with embedded scatterer
(a quantum dot) of a small size $w\ll L$ (see Fig.\ref{fig1}b). 
The quantum dot is characterized by 
the $2\times 2$ scattering matrix $\hat S$.
We are interested in a dc current arising in a ring under the slow
cyclic evolution of the scattering properties of a quantum dot.
We assume that there are no other effects which could generate 
circulating currents. In particular, (i) there is no magnetic flux through 
the ring;
(ii) the stationary scattering matrix $\hat S$ of the dot obeys
time reversal symmetry: $S_{12} = S_{21}$. 

We suppose that the scattering matrix 
$\hat S$ depends on a set of parameters $\{p_i\}$
which oscillate with frequency $\omega$:

\begin{equation}
\label{Eq5}
\begin{array}{c}
  \hat S = \hat S(p_1, p_2,\dots, p_{N_p}), \\
\ \\ 
  p_{i}(t) = p_{0i} + 2p_{1i}\cos(\omega t + \varphi_i), 
\end{array}
\end{equation}
\noindent
with $i = 1, 2, \dots, N_p .$
Then according to the Floquet theorem
one can write down the solution for 
the single-particle time-dependent Schr\"odinger equation
in a ring as follows \cite{MBring02}:

\begin{equation}
\label{Eq6}
\Psi_E(x,t) = e^{-iEt/\hbar}\sum\limits^\infty_{n=-\infty} 
e^{-in\omega t}  \left(
A_ne^{ik_n(x-L)} + B_ne^{-ik_nx}
\right).
\end{equation}

\noindent
Here the wave vector is $k_n = \sqrt{2m_eE_n/\hbar^2}$ 
with $Re[k_n]\geq  0$ and $Im[k_n]\geq 0$;
$E_n = E + n\hbar\omega$.
The Floquet eigenenergy $E$ is determined by the periodicity condition
and it is quantized like in the stationary ring problem
(see Ref.~\cite{MBring02} and below).
Each Floquet state $\Psi_E$ can be occupied by only one electron
(because of the Pauli principle)
and thus the wave function $\Psi_{E}$ must be normalized.

To find the circulating current $I_{dc}$ 
carried by the single-particle state $\Psi_E$ of interest here
we integrate the quantum mechanical current 
over the time period ${\cal{T}}= 2\pi/\omega$ and obtain

\begin{equation}
\label{Eq7}
I_{dc}^{(E)} = \sum\limits_{E_n > 0} \frac{e\hbar}{m_e}k_n
\left(
|A_n|^2 - |B_n|^2
\right).
\end{equation}

\noindent
We have restricted the summation over the propagating modes 
($E_{n}\equiv E + n\hbar\omega > 0$) only.

We are interested in the low frequency limit $\omega\to 0$.
To be more precise we assume 

\begin{equation}
\label{Eq8}
 \omega \ll \Delta^{(E)}/\hbar, \tau_{T}^{-1}.
\end{equation}

\noindent 
Here $\Delta^{(E)}$ is the level spacing close to 
the energy level $E$ under consideration.
The first inequality  $\hbar\omega\ll \Delta^{(E)}$
(characteristic for finite-size systems)
guarantees that the oscillating scatterer
does not produce interlevel transitions (Rabi oscillations).
Otherwise a large non-adiabatic current arises \cite{MBhidden}.
The second inequality $\omega\ll \tau_{T}^{-1}$
allows us to use an "instant scattering" approximation \cite{MB02}
which implies that scattering of electrons by 
the quantum dot is fast enough to ignore the change of the scattering
properties of a quantum dot during the particle traversal (reflection).
In this case the scattering properties of a quantum dot 
are completely described by the stationary scattering matrix 
$\hat S$ with parameters depending on time 
$\hat S(t) = \hat S(\{p_i(t)\})$ \cite{MBstrong02}. 
For instance, the Fourier coefficients $\hat S_{n\omega}$ of 
this scattering matrix define the amplitudes 
$\hat {\cal A}_n = \sqrt{{k}/{k_n}}\hat S_{n\omega}$ 
for scattering (transmission or reflection) 
of an electron with energy $E=\hbar^2 k^2/(2m_e)$ 
with the emission ($n<0$) 
or the absorption ($n>0$) of $n$ energy quanta $\hbar\omega$.

In the adiabatic limit \cite{Brouwer98, MBstrong02} 
knowledge of the solution of a scattering problem with 
small oscillating amplitudes is sufficient to calculate the pumped current 
in the lowest (first) order in $\omega$ at arbitrary oscillating strength (amplitudes).
Therefore, first, we consider the case when the parameters oscillate
with small amplitudes: 
$ p_{1i} \ll p_{0i}, \quad \forall~i$. 
We calculate the current in the lowest nonvanishing order 
in the oscillating amplitudes.
In this case it is enough to take into account
only the first sidebands \cite{MB02}. 
Thus in the expansion Eq.(\ref{Eq6}) 
we keep only the terms with $n = 0, \pm 1$ 
(we put all the coefficients $A_n, B_n$ for $|n|>1$ equal to zero).
The scattering matrix relates the incoming waves
$A_n, B_n$ to outgoing ones $A_ne^{-ik_nL}, B_ne^{-ik_nL}$.
We number the scattering channels as shown in Fig.~\ref{fig1}b.
Thus the scattering matrix defines the boundary conditions for 
an electron wave function Eq.(\ref{Eq6}) ($n=0, \pm 1$) as follows \cite{MBhidden}:

\begin{equation}
\label{Eq9}
\begin{array}{l}
A_ne^{-ik_nL} = \sum\limits_{m=0,\pm 1} \sqrt{\frac{k_{n-m}}{k_n}}  \\
~~~~~~~~~~\times\left( A_{n-m} S_{21,m\omega} 
+ B_{n-m} S_{22,m\omega} \right) \\
\ \\
B_ne^{-ik_nL} = \sum\limits_{m=0,\pm 1} \sqrt{\frac{k_{n-m}}{k_n}}  \\
~~~~~~~~~~\times\left( A_{n-m} S_{11,m\omega} 
+ B_{n-m} S_{12,m\omega} \right) .\\
\end{array}
\end{equation}

\noindent 
Note that on the RHS of the above equations for $n=\pm 1$ we have to put 
$A_{\pm 2} =0$ and $B_{\pm 2}=0$.
To obtain the current Eq.(\ref{Eq7}) to first
order in $\omega$ we expand $ e^{-ik_{\pm 1}L}$  in Eq.(\ref{Eq9}) as follows

\begin{equation}
\label{Eq10}
 e^{-ik_{\pm 1}L} \approx e^{-ikL}\left(
1 \mp i\frac{\omega}{\omega_0} 
- \frac{1}{2}\left(\frac{\omega}{\omega_0} \right)^2
\pm \frac{i}{6}\left(\frac{\omega}{\omega_0} \right)^3 \right),
\end{equation}

\noindent
where $\omega_0 = v/L$, and $v = \hbar k/m_e$ is an electron velocity.
In the above expansion we ignore all the terms containing additional
small factors $\omega/E$.

Solving Eq.(\ref{Eq9}) after a lengthy but rather straightforward calculation
we obtain the circulating current Eq.(\ref{Eq7})
(we restore the upper index $^{(l)}$)

\begin{equation}
\label{Eq11}
  I_{dc}^{(l)} = 
e\omega Im\left[\Gamma_{\omega}^{(l)}\theta_{-\omega}^{(l)} \right].
\end{equation}

\noindent Here we have introduced two real quantities.
The first one is characteristic of the spatial asymmetry of the
scatterer:
$ \theta = \frac{i}{2} ln(S_{11}/S_{22})$.
This quantity is real since $|S_{11}|^2 = |S_{22}|^2$. 
The second one is
$  \Gamma^{-1} = -i\left({e^{-iKL}}{S^{-1}_{12}} -1\right)$,  
where $K = k(\{p_{0i}\})$ is the solution of the dispersion equation for the stationary
problem (with $p_i = p_{01}$). This dispersion equation reads:
$Re\left[{e^{-ikL}}{S^{-1}_{12}} -1 \right] = 0$. 
From the dispersion equation it follows that the imaginary part of $\Gamma$ vanishes.
In particular, for the scattering matrix Eq.(\ref{Eq1}) we have
$K \equiv k^{(l)} = \frac{1}{L}[\pi l - (-1)^l \arcsin(\sqrt{T}) ]$ and 
$ \Gamma(k^{(l)}) = - (-1)^{l} \sqrt{{T}/{R}}$.

Equation (\ref{Eq11}) determines the current carried by the particular energy
level $E^{(l)}$. 
To find the full circulating current we have to sum Eq.(\ref{Eq11})
over all the occupied levels in the ring. 
Equation (\ref{Eq11}) shows that the adiabatically pumped current exists only if 
the time reversal symmetry (TRS) in the system is dynamically broken by
the oscillating scatterer. Such a breaking of TRS is a necessary
condition for the existence of an adiabatic pump effect in both open
\cite{MBstrong02} and closed systems. 
Otherwise if TRS is present then the Fourier coefficients for the real
quantities $\Gamma$ and $\theta$ are real and, thus, the current 
$I_{dc}^{(l)}$ Eq.(\ref{Eq11}) is identically zero.

Note that generally there is another 
necessary condition for the existence of an adiabatic quantum pump effect
for open and closed systems: The varying parameters 
must affect the spatial asymmetry of the scatterer \cite{MBstrong02}.
In our case the quantity $\theta$ [see Eq.(\ref{Eq1})] is a measure of
a spatial asymmetry of the scatterer.

Next we consider a large amplitude pump cycle.
To this end we apply the inverse Fourier transformation to the RHS 
of Eq.(\ref{Eq11}) and represent the circulating current as
an integral over the pump cycle (over the time period ${\cal T}=2\pi/\omega$):

\begin{equation}
\label{Eq12}
  I_{dc}^{(l)} 
= -\frac{e\omega}{4\pi} \int\limits_{0}^{\cal T} dt 
\Gamma^{(l)} \frac{\partial\theta^{(l)}}{\partial t}
= \frac{e\omega}{4\pi} \int\limits_{0}^{\cal T} dt 
\theta^{(l)} \frac{\partial\Gamma^{(l)}}{\partial t}.
\end{equation}

\noindent 
In equation (\ref{Eq12}) the integrand should be considered 
as a function of the time-dependent parameters $p_{i} = p_{i}(t)$
and the eigenenergy  $E^{(l)} = E^{(l)}(\{p_{i}(t)\})$
which adiabatically follows them.

The integral representation Eq.(\ref{Eq12}) 
allows us to calculate 
the circulating current for the pump cycle of an arbitrary strength. 
The only necessary condition is that 
the adiabaticity conditions Eq.(\ref{Eq8}) must hold at any point
of a pump cycle. Note that the level spacing 
depends on time since each eigenenergy is a function of time.
Therefore our consideration is valid if there is no level
crossing ($\Delta^{(E)} \neq 0$) during the pump cycle.

We would like to stress that 
the dual representation for the pumped current in Eq.(\ref{Eq12})
shows clearly that only true pump cycles Fig.\ref{fig2}b contribute to
the calculated quantity - the pumped current. This is in full agreement with
the Floquet scattering matrix approach to the pump effect in the open case \cite{MBstrong02}: 
The integral representation 
(see Eq.(18) in Ref.~\cite{MBstrong02} )
for the pumped current is a direct consequence of a differential representation
(see Eq.(17) in Ref.~\cite{MBstrong02} ). 
The later does not support a pseudo pump effect.

In conclusion, we have developed the scattering matrix approach 
to adiabatic quantum pumping in closed mesoscopic systems such as 
a ring with an embedded quantum dot. This formulation permits 
a direct comparison of pumping in open and closed systems. 
We have discussed the seemingly paradoxical nature of the result 
for open systems. Closer inspection of the two results demonstrates  
that the physics underlying the adiabatic quantum pump effect
in closed systems is very similar to that in open 
systems coupled to external reservoirs.The approach presented can be generalized 
to many channel rings and to closed systems with a more complicated topology. 
Experimental comparisons of pumping in open and closed systems would be very desirable,

\begin{acknowledgments}
This work is supported by the Swiss National Science Foundation.
\end{acknowledgments}


\begin{thebibliography}{11}


\bibitem{Thouless83}
    D.J. Thouless, Phys. Rev. B {\bf 27}, 6083 (1983).

\bibitem{SMCG99}
    M. Switkes, C. M. Marcus, K. Campman, and A. C. Gossard, 
    Science {\bf 283}, 1905 (1999).

\bibitem{Brouwer98}
    P. W. Brouwer, Phys. Rev. B {\bf 58}, R10135 (1998).

\bibitem{BTP94}
   M. B\"{u}ttiker, H. Thomas, and A. Pr\^{e}tre, 
   Z. Phys. B {\bf 94}, 133 (1994); M. B\"{u}ttiker,
   J. Phys. Condensed Matter {\bf 5}, 9361 (1993).

\bibitem{note} In practice the external impedance plays an important role
and can give rise to pumping via rectification 
(see Brouwer, Phys. Rev. B {\bf 63}, R121303 (2001) and L. DiCarlo, 
C. M. Marcus, and J. S. Harris, Jr., cond-mat/0304397). Here we assume 
for simplicity a zero-impedance external circuit. 

\bibitem{AA98}
   I.L. Aleiner and A.V. Andreev, Phys. Rev. Lett. {\bf 81}, 1286 (1998).

\bibitem{ZSA99}
    F. Zhou, B. Spivak, and B. Altshuler, 
    Phys. Rev. Lett. {\bf 82}, 608  (1999).

\bibitem{AK00}
    A. Andreev and A. Kamenev,
    Phys. Rev. Lett. {\bf 85}, 1294 (2000).

\bibitem{SAA00}
    T. A. Shutenko, I. L. Aleiner, and B. L. Altshuler, 
    Phys. Rev. B {\bf 61}, 10366 (2000).

\bibitem{WWG00}
    Y. Wei, J. Wang, and H. Guo, Phys. Rev. B {\bf 62}, 9947 (2000).

\bibitem{AEGS00}
    J. E. Avron, A. Elgart, G. M. Graf, and L. Sadun, 
    Phys. Rev. B {\bf 62}, 10618 (2000).
   
\bibitem{VAA01}
    M. G. Vavilov, V. Ambegaokar, and I. L. Aleiner,
    Phys. Rev. B {\bf 63}, 195313 (2001).

\bibitem{PB01}
    M. L. Polianski and P. W. Brouwer, 
    Phys. Rev. B {\bf 64}, 075304 (2001).

\bibitem{MM01}
    Y. Makhlin and A. Mirlin, Phys. Rev. Lett. {\bf 87}, 276803 (2001).

\bibitem{MB01}
    M. Moskalets and M. B\"{u}ttiker, 
    Phys. Rev. B. {\bf 64}, 201305 (2001).

\bibitem{AEGS01}
    J. E. Avron, A. Elgart, G. M. Graf, and L. Sadun,
    Phys. Rev. Lett. {\bf 87}, 236601 (2001);
    J. Math. Phys. 43: 3415 (2002). 

\bibitem{EWAL02}
    O. Entin-Wohlman, A. Aharony, and Y. Levinson, 
    Phys. Rev. B {\bf 65}, 195411 (2002).

\bibitem{MB02}
    M. Moskalets and M. B\"{u}ttiker, 
    Phys. Rev. B {\bf  66}, 035306  (2002).
  
\bibitem{Cohen02}
    D. Cohen, cond-mat/0208233 (unpublished).

\bibitem{MBstrong02}
    M. Moskalets and M. B\"{u}ttiker, Phys.Rev. B {\bf 66}, 205320 (2002).

\bibitem{SC02}
    P. Sharma and C. Chamon, cond-mat/0212201 (unpublished). 

\bibitem{MBhidden} 
    M. Moskalets and M. B\"{u}ttiker, cond-mat/0302586 (unpublished).

\bibitem{AEGS03}
    J.E. Avron, A. Elgart, G.M. Graf, and L. Sadum,   math-ph/0305049 (unpublished).

\bibitem{BDR03}
    A. Banerjee, S. Das and S. Rao,   cond-mat/0307324 (unpublished).

\bibitem{BTP93}
   M. B\"{u}ttiker, H. Thomas, and A. Pr\^{e}tre, 
   Phys. Lett. A {\bf 180}, 364 (1993).

\bibitem{PTB96}
   A. Pr\^{e}tre and H. Thomas, M. B\"{u}ttiker, 
   Phys. Rev. B {\bf 54}, 8130 (1996).

\bibitem{Friedel52}
   J. Friedel, Phil. Mag. {\bf 43}, 153 (1952).

\bibitem{MBring02} 
    M. Moskalets and M. B\"{u}ttiker, Phys.Rev. B {\bf 66}, 245321 (2002).


\end{thebibliography}
\end{document}